\documentclass[final,runningheads,a4paper]{llncs}

\usepackage[nomessages]{fp}
\usepackage{etoolbox}
\usepackage[utf8]{inputenc}

\makeatletter
\renewcommand\paragraph{\@startsection{paragraph}{4}{\z@}%
  {2.25ex \@plus 1ex \@minus .2ex}%
  {-0.75em}%
  {\normalfont\normalsize\bfseries}}
\makeatother

\usepackage[utf8]{inputenc}
\usepackage[english]{babel}

\usepackage{amsmath}
\usepackage{amsfonts}
\usepackage{amssymb}
\usepackage{amsbsy}

\usepackage{color}
\usepackage[usenames,dvipsnames]{xcolor}
\usepackage{mdframed}
\usepackage{tikz}
\usetikzlibrary{patterns}

\usepackage{listings}

\usepackage{stmaryrd}
\usepackage{graphicx}
\usepackage{marvosym}

\usepackage{pdfsync}
\usepackage{wasysym}

\usepackage{multirow}			
\usepackage{pgfplots}			
\usepgfplotslibrary{dateplot} 	
\usepackage{pgf-pie}

\usepackage{caption,subcaption}

\usepackage[official]{eurosym} 

\usepackage{fixltx2e} 
\usepackage{xspace} 

\usepackage{nicefrac}

\usepackage[inline,shortlabels]{enumitem} 
\newlist{inlinelist}{enumerate*}{1}
\setlist*[inlinelist,1]{%
	label=(\roman*),
}

\usepackage{xifthen}  

\usepackage{ifthen}

\usepackage[hyphens]{url}
\usepackage{cite}
\usepackage[hidelinks]{hyperref}

\hypersetup{
  breaklinks   = true,
  colorlinks   = true, 
  urlcolor     = blue, 
  linkcolor    = blue, 
  citecolor    = red   
}
\hypersetup{final} 

\usepackage{cleveref}

\usepackage[final,nomargin,inline,index]{fixme} 
\fxusetheme{color}

\FXRegisterAuthor{nico}{annicola}{\color{blue} {\underline{nico}}}
\FXRegisterAuthor{nicoq}{annicolaq}{\color{blue} {}}
\FXRegisterAuthor{bart}{anbart}{\color{magenta} {\underline{bart}}}
\FXRegisterAuthor{livio}{anlivio}{\color{red} {\underline{livio}}}
\FXRegisterAuthor{tizi}{antizi}{\color{orange} {\underline{tizi}}}

\lstdefinelanguage{coco}{
	commentstyle=\color{Gray},
	morecomment=[l]{//},
	morecomment=[s]{/*}{*/},
	classoffset=0,
	morekeywords={honesty,contract,process,system,if,then,else,
	nil,rec,specification,tellAndWait,send,receive,t,tellRetract,tellAndReturn},
	keywordstyle=\color{keyword}\bfseries,
	classoffset=1,
	morekeywords={tell,ask,do,t},
	keywordstyle=[1]\color{ForestGreen},
	classoffset=2,
	morekeywords={unit,int,session,string,boolean},
	keywordstyle=\color{NavyBlue},
	classoffset=0
}

\lstdefinelanguage{maude}{
	classoffset=0,
	morekeywords={mod,ops,eq,endm,red,quit,in},
	keywordstyle=\color{keyword}\bfseries,
	classoffset=1,
	morekeywords={tell,ask,do,t},
	keywordstyle=[1]\color{ForestGreen},
	classoffset=2,
	morekeywords={unit,int,session,string,boolean,exp},
	keywordstyle=\color{NavyBlue},
	classoffset=0
}

\lstdefinelanguage{java}{
	commentstyle=\color{Gray},
	morecomment=[l]{//},
	morecomment=[s]{/*}{*/},
	morestring=[b]",
    stringstyle=\color{NavyBlue},
	classoffset=0,
	morekeywords={public,private,static,final,class,extends,switch,case,break,finally,try,catch,void,int,boolean},
	keywordstyle=\color{keyword}\bfseries
}

\newcommand{\codefont}{\fontsize{10}{8}\selectfont}
\newcommand{\code}[1]{{\tt\codefont {#1}}}

\def\etc{\emph{etc}.\@\xspace}

\newcommand{\eg}{e.g.\@\xspace}
\newcommand{\ie}{i.e.\@\xspace}

\newcommand{\Real}[1]{\mathrm{Real}}

\def\cocoColor{\color{MidnightBlue}}
\newcommand{\cocoFmt}[1]{{\cocoColor{\code{#1}}}}

\let\greektau\tau
\renewcommand{\tau}{\cocoFmt{\greektau}}

\newcommand{\ifempty}[3]{%
  \ifthenelse{\isempty{#1}}{#2}{#3}%
}

\newcommand{\compile}[2]{\ifthenelse{\equal{#1}{yes}}{#2}{}}
\newcommand{\hidden}[1]{}

\let\greekgamma\gamma
\def\contrColor{\color{Plum}}
\newcommand{\contrFmt}[1]{{\contrColor{#1}}}

\renewcommand{\gamma}[1][]{\mathord{\contrFmt{\greekgamma}_{\contrFmt{#1}}}}

\crefname{appendix}{appendix}{appendices}
\Crefname{appendix}{Appendix}{Appendices}

\definecolor{LightGrey}{rgb}{0.95,0.95,0.95}
\definecolor{keyword}{HTML}{7F0055}

\newmdenv [linewidth=0pt]{mdNoFramed}
\makeatletter

\newcommand*{\tabminted@finalstrut}[1]{%
  \ifdim\prevdepth>0pt
    \ifdim\dp#1>\prevdepth
      \vskip\dimexpr(\dp#1)-\prevdepth\relax
    \fi
  \else
    \vskip\dimexpr(\dp#1)\relax
  \fi
}
\newcommand*{\@tabmintedend}{%
  \let\@finalstrut\tabminted@finalstrut
}
\makeatother

\newcommand{\opreturn}{OP\_RETURN\xspace}

\definecolor{rosso}{RGB}{220,57,18}
\definecolor{giallo}{RGB}{255,153,0}
\definecolor{blu}{RGB}{102,140,217}
\definecolor{verde}{RGB}{16,150,24}
\definecolor{viola}{RGB}{153,0,153}
\definecolor{nero}{RGB}{0,0,0}

\makeatletter

\tikzstyle{chart}=[
legend label/.style={font={\scriptsize},anchor=west,align=left},
legend box/.style={rectangle, draw, minimum size=5pt},
axis/.style={black,semithick,->},
axis label/.style={anchor=east,font={\tiny}},
]

\tikzstyle{bar chart}=[
chart,
bar width/.code={
	\pgfmathparse{##1/2}
	\global\let\bar@w\pgfmathresult
},
bar/.style={very thick, draw=white},
bar label/.style={font={\bf\small},anchor=north},
bar value/.style={font={\footnotesize}},
bar width=.75,
]

\tikzstyle{pie chart}=[
chart,
slice/.style={line cap=round, line join=round, very thick,draw=white},
pie title/.style={font={\bf}},
slice type/.style 2 args={
	##1/.style={fill=##2},
	values of ##1/.style={}
}
]

\pgfdeclarelayer{background}
\pgfdeclarelayer{foreground}
\pgfsetlayers{background,main,foreground}

\newcommand{\mypie}[3][]{
	\begin{scope}[#1]
		\pgfmathsetmacro{\curA}{90}
		\pgfmathsetmacro{\r}{1}
		\def\c{(0,0)}
		\node[pie title] at (90:1.3) {#2};
		\foreach \v/\s in{#3}{
			\pgfmathsetmacro{\deltaA}{\v/100*360}
			\pgfmathsetmacro{\nextA}{\curA + \deltaA}
			\pgfmathsetmacro{\midA}{(\curA+\nextA)/2}
			
			\path[slice,\s] \c
			-- +(\curA:\r)
			arc (\curA:\nextA:\r)
			-- cycle;
			\pgfmathsetmacro{\d}{max((\deltaA * -(.5/50) + 1) , .5)}
			
			\begin{pgfonlayer}{foreground}
				\path \c -- node[pos=\d,pie values,values of \s]{$\v\%$} +(\midA:\r);
			\end{pgfonlayer}
			
			\global\let\curA\nextA
		}
	\end{scope}
}

\newcommand{\statsdate}{January 1st, 2017\xspace}
\newcommand{\ethscandate}{January 1st, 2017\xspace}
\newcommand{\btcscandate}{January 1st 2017\xspace}

\newcommand{\btcsize}{96 GB}
\newcommand{\btctx}{184,045,240}
\newcommand{\btcvolume}{83,178}
\newcommand{\btcmarketcap}{15,482}

\newcounter{valbtccontracts}
\newcommand{\btccontracts}{\arabic{valbtccontracts}\xspace}

\newcounter{valethcontracts}
\newcommand{\ethcontracts}{\arabic{valethcontracts}\xspace}

\newcounter{valtotcontracts}
\newcommand{\totcontracts}{\arabic{valtotcontracts}\xspace}

\newcommand{\mytitle}{An empirical analysis of smart contracts: platforms, applications, and design patterns}

\begin{document}

\mainmatter

\titlerunning{An empirical analysis of smart contracts}
\title{\mytitle}

\author{Massimo Bartoletti \and Livio Pompianu}
\authorrunning{Bartoletti M., Pompianu, L.}
\tocauthor{Massimo Bartoletti and Livio Pompianu}
\institute{Universit\`a degli Studi di Cagliari, Cagliari, Italy \\ \email{\{bart,livio.pompianu\}@unica.it}}

\maketitle

\begin{abstract}
  Smart contracts are computer programs that can be consistently 
  executed by a network of mutually distrusting nodes, 
  without the arbitration of a trusted authority.
  Because of their resilience to tampering,
  smart contracts are appealing in many scenarios, 
  especially in those which require transfers of money 
  to respect certain agreed rules
  (like in financial services and in games).
  Over the last few years many platforms for smart contracts have been proposed,
  and some of them have been actually implemented and used.
  We study how the notion of smart contract is interpreted 
  in some of these platforms.
  Focussing on the two most widespread ones, Bitcoin and Ethereum,
  we quantify the usage of smart contracts
  in relation to their application domain.
  We also analyse the most common programming patterns
  in Ethereum, where the source code of smart contracts is available.
\end{abstract}

\section{Introduction}
\label{Introduction}

Since the release of Bitcoin in 2009~\cite{bitcoin}, 
the idea of exploiting its enabling technology 
to develop applications beyond currency has been receiving 
increasing attention~\cite{Bonneau15ieeesp}.
In particular,
the public and append-only ledger of transaction (the \emph{blockchain})
and the decentralized consensus protocol that Bitcoin nodes use to extend it,
have revived Nick Szabo's idea of \emph{smart contracts}
--- \ie programs whose correct execution is automatically enforced without relying on a trusted authority~\cite{Szabo97firstmonday}.
The archetypal implementation of smart contracts is Ethereum~\cite{ethereum}, 
a platform where they are rendered in a Turing-complete language.
The consensus protocol of Ethereum ensures that all and only the 
valid updates to the contract states are recorded on the blockchain,
so ensuring their correct execution.

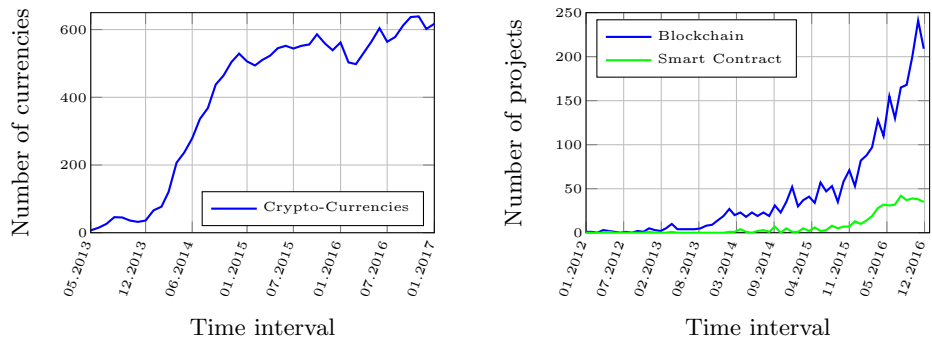
\begin{figure}[t]
  \hspace{-5pt}
  \begin{tabular}{cc}
    \begin{subfigure}[b]{0.5\textwidth}
		\begin{tikzpicture}
		\begin{axis}[
		width  = 1\linewidth,
		height = 4.5cm,
		date coordinates in=x, date ZERO=2013-05-01,
		xmin=2013-05-01, xmax=2017-01-01,
		xtick={2013-05-01, 2013-12-01, 2014-06-01, 2015-01-01, 2015-07-01, 2016-01-01, 2016-07-01, 2017-01-01},
		ymin=0, ymax=650,
		xmajorgrids = true,
		ymajorgrids = true,
		xticklabel style={rotate=70, anchor=east, xticklabel}, xticklabel=\tiny\month.\year,
		yticklabel style={font=\tiny,xshift=0.5ex},
		legend pos=south east, legend style={font=\tiny}, legend cell align=left,
		xlabel absolute, xlabel style={yshift=-0.5cm}, xlabel={Time interval},
		ylabel absolute, ylabel style={yshift=-0.3cm}, ylabel={Number of currencies}
		]
		\pgfplotstableread[col sep=comma]{results/Coinmarketcap.csv}\data
		\addplot [color=blue, thick] table[x index = {0}, y index = {1}]{\data};
		\legend{Crypto-Currencies}
		\end{axis}
		\end{tikzpicture}
    \end{subfigure}
    \hspace{5pt}
    &
      \begin{subfigure}[b]{0.5\textwidth}
		\begin{tikzpicture}
		\begin{axis}[
		width  = 1\linewidth,
		height = 4.5cm,
		date coordinates in=x, date ZERO=2012-01-01,
		xmin=2012-01-01, xmax=2017-01-01,
		ymin=0, ymax=250,
		xmajorgrids = true,
		ymajorgrids = true,
		try min ticks=8,
		xticklabel style={rotate=70, anchor=east, xticklabel}, xticklabel=\tiny\month.\year,
        yticklabel style={font=\tiny,xshift=0.5ex},
        legend pos=north west, legend style={font=\tiny}, legend cell align=left,
		xlabel absolute, xlabel style={yshift=-0.5cm}, xlabel={Time interval},
		ylabel absolute, ylabel style={yshift=-0.3cm}, ylabel={Number of projects}
		]
		\pgfplotstableread[col sep=comma]{results/Github.csv}\data
		\addplot [color=blue, thick] table[x index = {0}, y index = {1}]{\data};
		\pgfplotstableread[col sep=comma]{results/Github.csv}\data
		\addplot [color=green, thick] table[x index = {0}, y index = {2}]{\data};
		\legend{Blockchain,Smart Contract}
		\end{axis}
		\end{tikzpicture}
      \end{subfigure}
      \hspace{40pt}
  \end{tabular}
  \caption{On the left, monthly trend of the number of crypto-Currencies hosted on \href{http://coinmarketcap.com/}{coinmarketcap.com}. 
    On the right, number of new projects related to blockchains and smart contracts which are created every month on \href{http://github.com}{github.com}.}
  \label{fig:StatisticsIntro}
\end{figure}

Besides Bitcoin and Ethereum, a remarkable number of alternative platforms
have flourished over the last few years,
either implementing crypto-currencies or some forms of 
smart contracts~\cite{Clack16corr,Luu16ccs,MultiChainGoodBadLazy,BitcoinContract,MakingSenseContracts}.
For instance, the number of crypto-currencies 
hosted on~\href{http://coinmarketcap.com/}{coinmarketcap.com}
has increased from 0 to more than 600 since 2012;
the number of \href{http://github.com}{github} projects related to blockchains and smart contracts
has reached, respectively, $2,715$ and $445$ units (see~\Cref{fig:StatisticsIntro}).
In the meanwhile, ICT companies and some national governments have started 
dealing with these topics~\cite{UK16report, Japan16report}, 
also with significant investments.

Despite the growing hype on blockchains and smart contracts,
the understanding of the actual benefits of these technologies,
and of their trustworthiness and security, has still to be assessed.
In particular, 
the consequences of unsafe design choices for the programming languages for smart contracts
can be fatal, as witnessed by the unfortunate epilogue of the DAO contract~\cite{DAO}, 
a crowdfunding service plundered of $\sim50M$ USD because of a programming error.
Since then, many other vulnerabilities in smart contract have been reported~\cite{ABC17post,Luu16ccs,VitalikThinkingSecurity,ENSanotherbug}.

Understanding how smart contracts are used and how they are implemented
could help designers of smart contract platforms
to create new domain-specific languages 
(not necessarily Turing complete~\cite{Brown16corda,Churyumov16byteball,Frantz16ecas,Popejoy16kadena}),
which \emph{by-design} avoid vulnerabilities as the ones discussed above.
Further, this knowledge could help to improve analysis techniques for smart contracts
(like \eg the ones in~\cite{Bhargavan16solidether,Luu16ccs}),
by targeting contracts with specific programming patterns.

  \paragraph{Contributions.}

This paper is a methodic survey on smart contracts, 
with a focus on Bitcoin and Ethereum --- the two most widespread platforms currently supporting them.
Our main contributions can be summarised as follows:
\begin{itemize}

\item in \Cref{sec:platforms} we examine the Web for news about smart contracts
  in the period from June 2013 to September 2016, 
  collecting data about 12 platforms.
  We choose from them a sample of 6 platforms which are amenable to analytical investigation.
  We analyse and compare several aspects of the platforms in this sample, 
  mainly concerning their usage, and their support for programming  smart contracts.
  
\item in \Cref{sec:smartcontracts} 
  we propose a taxonomy of smart contracts, 
  sorting them into categories which reflect their application domain.
  We collect from the blockchains of Bitcoin and Ethereum
  a sample of 834 smart contracts, 
  which we classify according to our taxonomy. 
  We then study the usage of smart contracts,
  measuring the distribution of their transactions by category.
  This allows us to compare the different usage of Bitcoin and Ethereum
  as platforms for smart contracts.

\item in \Cref{sec:DesignPatterns} we consider the source code of the Ethereum contracts in our sample.
  We identify 9 common design patterns,
  and we quantify their usage in contracts, 
  also in relation to the associated category.
  Together with the previous point, ours constitutes the first quantitative 
  investigation on the usage and programming of smart contract in Ethereum.

\end{itemize}

\noindent
All the data collected by our survey are availble online at: \href{https://goo.gl/pOswL8}{\code{goo.gl/pOswL8}}.

\section{Platforms for smart contracts}
\label{sec:platforms}

In this~\namecref{sec:platforms} we analyse
various platforms for smart contracts.  
We start by presenting the methodology we have followed to choose the
candidate platforms (\Cref{subsection:platfrommethodology}). 
Then we describe the key features of each platform, 
pinpointing differences and similarities,
and drawing some general statistics (\Cref{subsection:platformanalysis}).

  \subsection{Methodology}
\label{subsection:platfrommethodology}

To choose the platforms subject of our study, 
we have drawn up a candidate list by examining
all the articles of 
\href{http://www.coindesk.com}{\code{coindesk.com}}
in the ``smart contracts'' category%
\footnote{\url{http://www.coindesk.com/category/technology/smart-contracts-news}}.
Starting from June 2013, when the first article appeared, 
up to the 15th of September 2016, 
175 articles were published, 
describing projects, events, companies and technologies related 
to smart contracts and blockchains.
By manually inspecting all these articles, 
we have found references to 12 platforms:
Bitcoin, Codius, Counterparty, DAML, Dogeparty, Ethereum, Lisk, Monax, Rootstock, Symbiont, Stellar, and Tezos.

We have then excluded from our sample the platforms which, 
at the time of writing, 
do not satisfy one of the following criteria:
\begin{inlinelist}
\item \label{item:platform-methodology:launched} 
  have already been launched,
\item \label{item:platform-methodology:running}
  are running and supported from a community of developers, and
\item \label{item:platform-methodology:accessible}
  are publicly accessible. 
\end{inlinelist}
For the last point we mean that, \eg, 
it must be possible to write a contract and test it, 
or to explore the blockchain through some tools, 
or to run a node. 
We have inspected each of the candidate platforms, 
examining the related resources available online 
(\eg, official websites, white-papers, forum discussions, \etc)
After this phase, we have removed 6 platforms from our list: 
Tezos and Rootstock, as they do not satisfy condition~\ref{item:platform-methodology:launched};
Codius and Dogeparty, which violate condition~\ref{item:platform-methodology:running},
DAML and Symbiont, which violate~\ref{item:platform-methodology:accessible}.
Summing up, we have a sample of 6 platforms: 
Bitcoin, Ethereum, Counterparty, Stellar, Monax and Lisk, 
which we discuss in the following.

  \subsection{Analysis of platforms}
\label{subsection:platformanalysis} 

We now describe the general features of the collected platforms, 
focussing on:
\begin{inlinelist}

\item whether the platform has its own blockchain, 
  or if it just piggy-backs on an already existing one;

\item for platforms with a public blockchain,
  their consensus protocol, and whether the blockchain is 
  public or private to a specific set of nodes;

\item the languages used to write  smart contracts.

\end{inlinelist}

\paragraph{Bitcoin} \cite{bitcoin} is a platform for transferring
digital currency, the bitcoins (BTC).  It has been the first
decentralized cryptocurrency to be created, and now is the one with
the largest market capitalization. %
The platform relies on a public blockchain to record the complete
history of currency transactions.  
The nodes of the Bitcoin network
use a consensus algorithm based 
moderately hard \emph{``proof-of-work''} puzzles to establish
how to append a new block of transactions to the blockchain.  
Nodes work in competition to generate the next block of the chain. 
The first node that solves the puzzle earns a reward in BTC.

Although the main goal of Bitcoin is to transfer currency, 
the immutability and openness of its blockchain have inspired the
development of protocols that implement (limited forms of) smart contracts.  
Bitcoin features a non-Turing complete scripting language, 
which allows to specify under which conditions
a transaction can be redeemed.
The scripting language is quite limited,
as it only features some basic arithmetic, logical, and crypto operations
(\eg, hashing and verification of digital signatures).
A further limitation to its expressiveness is the fact that only
a small fraction of the nodes of the Bitcoin network
processes transactions whose script is more complex than verifying 
a signature%
\footnote{As far as we know, currently only the \emph{Eligius} mining pool accepts more general transactions
(called \emph{non-standard} in the Bitcoin community).
However, this pool only mines \mbox{$\sim 1\%$} of the total mined blocks~\cite{Banasik16esorics}.}.

\paragraph{Ethereum}\cite{ethereum} 
is the second platform for market capitalization, after Bitcoin.
Similarly to Bitcoin, it relies on a public
blockchain, with a consensus algorithm similar to that of Bitcoin%
\footnote{The consensus mechanism of Ethereum is a variant of the 
GHOST protocol in~\cite{Sompolinsky15fc}.
}.
Ethereum has its own currency, caller \emph{ether} (ETH).
Smart contracts are written in a stack-based bytecode language~\cite{ethereumyellowpaper},
which is Turing-complete, unlike Bitcoin's.
There also exist a few high level languages 
(the most prominent being \emph{Solidity}%
\footnote{Solidity: \url{http://solidity.readthedocs.io/en/develop/index.html}}%
), 
which compile into the bytecode language.
Users create contracts and invoke their functions
by sending transactions to the blockchain,
whose effects are validated by the network.
Both users and contracts can store money and send/receive ETH 
to other contracts or users.

\paragraph{Counterparty}\cite{counterparty}
is a platform without its own blockchain; 
rather, it embeds its data into Bitcoin transactions.
While the nodes of the Bitcoin network 
ignore the data embedded in these transactions, 
the nodes of Counterparty recognise and interpret them. 
Smart contracts can be written in the same language used by Ethereum.
However, unlike Ethereum,
no consensus protocol is used to validate the results of computations%
\footnote{See FAQ: How do Smart Contracts ``form a consensus'' on \mbox{Counterparty}? \url{http://counterparty.io/docs/faq-smartcontracts/\#how-do-smart-contracts-form-a-consensus-on-counterparty}}.
Counterparty has its own currency, which can be transferred between users,
and be spent for executing contracts.
Unlike Ethereum, nodes do not obtain fees for executing contracts;
rather, the fees paid by clients are destroyed,
and nodes are indirectly rewarded from the inflation of the currency.
This mechanism is called \emph{proof-of-burn}.

\paragraph{Stellar}\cite{Stellar} 
features a public blockchain with its own cryptocurrency, 
governed by a consensus algorithm inspired to
federated Byzantine agreement~\cite{SCP}.  
Basically, a node agrees on a transaction
if the nodes in its neighbourhood 
(that are considered more trusted than the others) agree as well.
When the transaction has been accepted by enough nodes of the network,
it becomes unfeasible for an attacker to roll it back, 
and it is considered as confirmed.
Compared to \emph{proof-of-work}, this protocol consumes 
far less computing power, since it does not involve solve cryptographic puzzles.
Unlike Ethereum, there is no specific language for smart contracts; 
still, it is possible to gather together some transactions
(possibly ordered in a chain) and write them atomically in the blockchain.
Since transactions in a chain can involve different addresses,
this feature can be used to implement basic smart contracts. 
For instance, assume that a participant
$A$ wants to pay $B$ only if $B$ promises to pay $C$ after receiving the
payment from $A$.
This behaviour can be enforced by putting these transactions in the same chain.
While this specific example can be implemented on Bitcoin as well, 
Stellar also allows to batch operations different from payments%
\footnote{\url{https://www.stellar.org/developers/guides/concepts/operations.html}},
\eg creating new accounts. 
Stellar features special accounts, called \emph {multisignature},
which can be handled by several owners.
To perform operations on these accounts, 
a threshold of consensus must be reached among the owners.
Transaction chaining and multisignature accounts can be
combined to create more complex contracts.

\paragraph{Monax} \cite{Monax} %
supports the execution of Ethereum contracts,
without having its own currency.
Monax allows users to create private blockchains,
and to define authorisation policies for accessing them.
Its consensus protol%
\footnote{\url{https://tendermint.com/}} 
is organised in rounds, 
where a participant proposes a new block of transactions, 
and the others vote for it.  
When a block fails to be approved, 
the protocol moves to the next round, 
where another participant will be in charge of proposing blocks. 
A block is confirmed when it is approved by
at least 2/3 of the total voting power.
%

\paragraph{Lisk} \cite{Lisk}
has its own currency, and a public blockchain with a 
\emph{delegated proof-of-stake} consensus mechanism%
\footnote{\url{https://lisk.io/documentation?i=lisk-handbooks/DelegateHandbook}}.
More specifically, 101 active delegates, 
each one elected by the stakeholders, 
have the authority to generate blocks. 
Stakeholders can take part to the electoral process, 
by placing votes for delegates in their favour, 
or by becoming candidates themselves.
Lisk supports the execution of Turing-complete smart contracts,
written either in JavaScript or in Node.js.
Unlike Ethereum, determinism of executions is not ensured by the language:
rather, programmers must take care of it, 
\eg by not using functions like \textit{Math.random}.
Although Lisk has a main blockchain, 
each smart contract is executed on a separated one.
Users can deposit or withdraw currency
from a contract to the main chain, while avoiding double spending. 
Contract owners can customise their blockchain before deploying their contracts,
\eg choosing which nodes can participate to the consensus mechanism.

\begin{table}[t]
  \begin{center}
    \hspace*{0em}
    \small
    \scalebox{0.75}{%
      \begin{tabular}{|c|c|c|c|c|c|c|c|}
        \hline
        \multirow{2}{4.5em}{\textbf{Platform}} & \multicolumn{3}{|c|}{\textbf{Blockchain}} & \multirow{2}{9.7em}{\textbf{Contract Language}} & \multirow{2}{5.5em}{\textbf{		Total Tx}} & \textbf{Volume} & \textbf{Marketcap} \\
        \cline{2-4}
                                               & \textbf{Type} & \textbf{Size} & \textbf{Block int.} & & & (K USD) & (M USD) \\
        \hline\hline
        \textbf{Bitcoin} & \multirow{2}{2.9em}{Public} & \multirow{2}{2.8em}{\btcsize} & \multirow{2}{3.3em}{10 min.} & Bitcoin scripts + signatures & \btctx & \btcvolume & \btcmarketcap \\ 
        \cline{1-1} \cline{5-8}
        \textbf{Counterparty} & & & & EVM bytecode & 12,170,386 & 33 & 4 \\ 
        \hline
        \textbf{Ethereum} & Public & 17-60 GB & 12 sec. & EVM bytecode & 14,754,984 & 10,354 & 723 \\ 
        \hline
        \textbf{Stellar} & Public & ? & 3 sec. & Transaction chains + signatures & ? & 35 & 17 \\
        \hline
        \textbf{Monax} & Private & ? & Custom & EVM bytecode + permissions & ? & n/a & n/a \\
        \hline
        \textbf{Lisk} & Private & ? & Custom & JavaScript & ? & 45 & 15 \\
        \hline
      \end{tabular}
    } 
    \caption{General statistics of platforms for smart contracts.}
    \label{table:platforms}
  \end{center}
\end{table}

\medskip

\Cref{table:platforms} summarizes the main features of the analysed platforms. 
The question mark in some of the cells indicates that we
were unable to retrieve the information
(\eg, we have not been able to determine the size of Monax blockchains,
since they are private). 
The first three columns next to the platform name describe features of the blockchain: 
whether it is public; its size; the average time between two consecutive blocks. 
Note that Bitcoin and Counterparty share the same cell, 
since the second platform uses the Bitcoin blockchain.  
Measuring the size of the Ethereum blockchain depends on which client and which
pruning mode is used. 
For instance, using the \href{https://github.com/ethereum/go-ethereum/wiki/geth}{Geth} client, 
we obtain a measure of 17GB in ``fast sync'' mode, and of 60GB in ``archive'' mode%
\footnote{\url{https://redd.it/5om2lw}}.
In platforms with private blockchains, 
their block interval is custom.
The fifth column describes the support for writing contracts.
The sixth column shows the total number of transactions%
\footnote{Sources: 
  \url{https://blockchain.info/charts/n-transactions-total} (for Bitcoin),
  \url{https://blockscan.com} (Counterparty), and
  \url{https://etherscan.io} (Ethereum).}.
The last two columns show the daily volume of currency tranfers,
and the market capitalisation of the currency 
(both in USD, rounded, respectively, to thousands and millions)%
\footnote{Market capitalization estimated by~\url{http://coinmarketcap.com}.}.
All values reported on \Cref{table:platforms} are updated to~\statsdate.

\section{Analysing the usage of smart contracts}
\label{sec:smartcontracts}

In this~\namecref{sec:smartcontracts} we analyse the usage of smart contracts,
proposing a classification which reflects their application domain.
Then, focussing on Bitcoin and Ethereum,
we quantify the usage of smart contracts
in relation to their application domain.
We start by presenting the methodology we have followed to 
sample and classify Bitcoin and Ethereum smart contracts
(\Cref{subsection:smartcontractmethodology}). 
Then, we introduce our classification and our statistical analysis 
(\Cref{sec:smartContracts:taxonomy,sec:smartContracts:quantitative}).

  \subsection{Methodology}
\label{subsection:smartcontractmethodology}

We sample contracts from Bitcoin and Ethereum as follows:
\begin{itemize}

\item for Ethereum, we collect on \ethscandate
  all the contracts marked as ``verified''
  on the blockchain explorer \href{https://etherscan.io/contractsVerified}{\code{etherscan.io}}.
  This means that the contract bytecode stored on the blockchain 
  matches the source code %
  (generally written in a high level language, such as Solidity) submitted to the explorer.
  In this way, we obtain a sample of \ethcontracts contracts.

\item for Bitcoin, we start by observing that 
  many smart contracts save their metadata on the blockchain 
  through the \opreturn
  instruction of the Bitcoin scripting language~\cite{BitcoinContract,MakingSenseContracts,BitcoinOpReturnDescription,Bartoletti16opreturn}.
  We then scan the Bitcoin blockchain on \btcscandate,
  searching for transactions that embed in an \opreturn
  some metadata attributable to a Bitcoin smart contract.
  To this purpose we use an explorer%
  \footnote{\url{https://github.com/BitcoinOpReturn/OpReturnTool}}
  which recognises \btccontracts smart contracts,
  and extracts all the transactions related to them.
\end{itemize}

  \subsection{A taxonomy of smart contracts}
\label{sec:smartContracts:taxonomy}

We propose a taxonomy of smart contracts into five categories,
which describe their intended application domain.  
We then classify the contracts in our sample according to the taxonomy.
To this purpose, for Ethereum contracts  we manually inspect the
Solidity source code, while for Bitcoin contracts we search their
web pages and related discussion forums.
After this manual investigation,
we distribute all the contracts into the five categories,
that we present below.

\begin{description}
\item[Financial.]  Contracts in this category 
  manage, gather, or distribute money as preeminent feature.
  Some contracts certify the ownership of a
  real-world asset, endorse its value, and keep track of trades 
  (\eg, \href{http://coloredcoins.org/explorer/}{Colu}
  currently tracks over 50,000 assets on Bitcoin).
  Other contracts implement crowdfunding services, 
  gathering money from investors in order to fund projects
  (the Ethereum \href{https://forum.daohub.org/}{DAO} project was 
  the most representative one, until its collapse due to an attack in June 2016).
  High-yield investment programs are a type of Ponzi schemes~\cite{Bartoletti17ponzi} 
  that collect money from users under the promise that they will
  receive back their funds with interest if new
  investors join the scheme
  (\eg, \href{http://governmental.github.io/GovernMental/}{Government},
  \href{https://www.kingoftheether.com/}{KingOfTheEtherThrone}).
  Some contracts provide an insurance on setbacks which are digitally provable
  (\eg, \href{https://fdi.etherisc.com/}{Etherisc} sells
  insurance policies for flights; if a flight is delayed or cancelled, one obtains a refund).
    Other contracts publish advertisement messages
    (\eg, \href{http://pixelmap.io/}{PixelMap} 
    is inspired to the \href{https://en.wikipedia.org/wiki/The_Million_Dollar_Homepage}{Million Dollar Homepage}).
  
\item[Notary.] Contracts in this category exploit the immutability of the blockchain
  to store some data persistently,
  and in some cases to certify their ownership and provenance.
  Some contracts allow users to write the hash of a document on the blockchain, 
  so that they can prove document existence and integrity
  (\eg, \href{https://proofofexistence.com/}{Proof of Existence}). 
  Others allow to declare copyrights on digital arts files, 
  like photos or music (\eg, \href{https://monegraph.com/}{Monegraph}).
  Some contracts (\eg, \href{https://eternitywall.it/} {Eternity Wall}) 
  just allow users to write down on the blockchain messages that everyone can read.
  Other contracts associate users to addresses (often represented as public keys),
  in order to certify their identity
  (\eg, \href{https://proofofphysicaladdress.com/}{Physical Address}).

\item[Game.] This category gathers contracts which implement 
  \emph{games of chance}
  (\eg, \href{https://etherscan.io/address/0x2ef76694fBfD691141d83F921A5ba710525De9B0#code}{LooneyLottery},
  \href{https://etherscan.io/address/0x2AB9f67A27f606272189b307052694D3a2B158bA#code}{Dice},
  \href{https://etherscan.io/address/0x18a672e11d637fffadccc99b152f4895da069601#code}{Roulette},
  \href{https://etherscan.io/address/0x1d77340D3819007BbfD7fdD37C22BD3b5c311350#code}{RockPaperScissors})
  and \emph{games of skill}
  (\eg, \href{http://www.bspend.com/etherization}{Etherization}),
  as well as some games which mix chance and skill
  (\eg, \href{https://etherscan.io/address/0x4ed65e408439a7f6459b5cfbd364f373bd6ed5f7#comments}{PRNG challenge} pays for the solution of a puzzle).

\item[Wallet.] The contracts in this category 
  handle keys, send transactions, manage money, deploy and watch contracts, 
  in order to simplify the interaction with the blockchain. 
  Wallets can be managed by one or many owners, in the latter case
  requiring multiple authorizations
  (like, \eg in \href{https://etherscan.io/address/0xA2D4035389aae620E36Bd828144b2015564C2702#code}{Multi-owned}).

\item[Library.] These contracts implement general-purpose operations
  (like \eg, math and string transformations),
  to be used by other  contracts.  

\end{description}

  \subsection{Quantifying the usage of smart contracts by category}
\label{sec:smartContracts:quantitative}

\begin{table}[t]
  \begin{center}
    \small
    \scalebox{0.75}{%
      \begin{tabular}{|c|c|c|c|}
      	\hline
      	           \textbf{Category}             & \textbf{Platform} & \textbf{Contracts} & \textbf{Transactions} \\ \hline\hline
      	 \multirow{2}{8em}{\textbf{Financial}}   &      Bitcoin      &         6          &        470,391        \\ \cline{2-4}
      	                                         &     Ethereum      &        373         &        624,046        \\ \hline
      	   \multirow{2}{8em}{\textbf{Notary}}    &      Bitcoin      &         17         &        443,269        \\ \cline{2-4}
      	                                         &     Ethereum      &         79         &        35,253         \\ \hline
      	    \multirow{2}{8em}{\textbf{Game}}     &      Bitcoin      &         0          &           0           \\ \cline{2-4}
      	                                         &     Ethereum      &        158         &        58,257         \\ \hline
      	   \multirow{2}{8em}{\textbf{Wallet}}    &      Bitcoin      &         0          &           0           \\ \cline{2-4}
      	                                         &     Ethereum      &         17         &         1,342         \\ \hline
      	  \multirow{2}{8em}{\textbf{Library}}    &      Bitcoin      &         0          &           0           \\ \cline{2-4}
      	                                         &     Ethereum      &         29         &        37,034         \\ \hline\hline
      	\multirow{2}{8em}{\textbf{Unclassified}} &      Bitcoin      &         0          &           0           \\ \cline{2-4}
      	                                         &     Ethereum      &        155         &         3,679         \\ \hline\hline
      	   \multirow{3}{8em}{\textbf{Total}}     &      Bitcoin      & \btccontracts      &        913,660        \\ \cline{2-4}
      	                                         &     Ethereum      & \ethcontracts      &        759,611        \\ \cline{2-4}
      	                                         &      Overall      & \totcontracts      &       1,673,271       \\ \hline
      \end{tabular}
    } 
  \end{center}
  \caption{Transactions by category.}
  \label{fig:smartcontracts}
\end{table}

\begin{figure}
  \centering
	\begin{tikzpicture}
	\begin{axis} 
		[ybar,
		enlargelimits=0.15,
		width  = 13cm,
		height = 5.5cm,
		legend style={at={(0.75,0.75)},anchor=south},
		scaled y ticks = false,
		legend columns=3,
		symbolic x coords={Financial,Notary,Wallet,Game,Library,Unclassified},
		xlabel absolute, xlabel style={yshift=+0.2cm},
		ymin=0, ymax=80,
                xtick pos=left,
                ytick pos=left,
		xtick={Financial,Notary,Wallet,Game,Library,Unclassified}]
	
		\addplot
		[fill=blue!50]
		coordinates
		{(Financial,51.5) (Notary,48.5) (Game,0) (Wallet,0) (Library,0) (Unclassified,0)};
		
		\addplot
		[fill=red!50]
		coordinates
		{(Financial,82.2) (Notary,4.6) (Game,7.7) (Wallet,0.2) (Library,4.9) (Unclassified,0.5)};
		
		\addplot
		[fill=violet!70]
		coordinates
		{(Financial,65.5) (Notary,28.6) (Game,3.5) (Wallet,0.1) (Library,2.2) (Unclassified,0.2)};
		
		\legend{Bitcoin, Ethereum, Overall}
		\end{axis}
	\end{tikzpicture}
	\caption{Distribution of transactions by category.}
	\label{fig:TransactionsAndLength}
\end{figure}
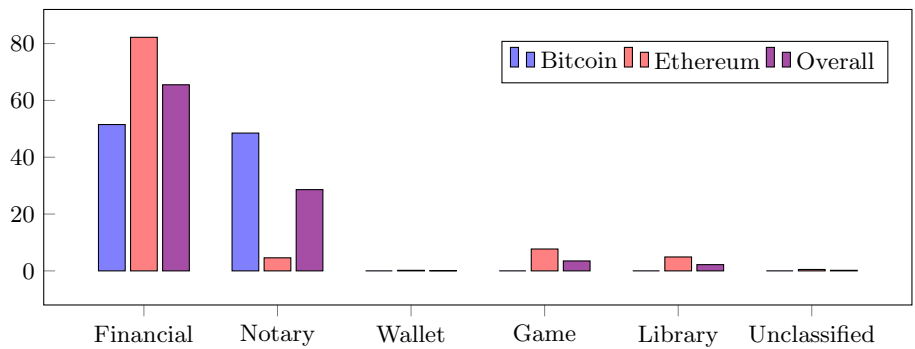

We analyse all the transactions related to the \totcontracts smart contracts
in our sample.
\Cref{fig:smartcontracts} displays how the transactions are
distributed in the categories of~\Cref{sec:smartContracts:taxonomy}.  
For both Bitcoin and Ethereum, we show
the number of detected contracts (third column), and
the total number of transactions (fourth column).

Overall, we have 1,673,271 transactions. 
Notably, although Bitcoin contracts are fewer than those running on Ethereum,
they have a larger amount of transactions each.
A clear example of this is witnessed by the financial category,
where 6 Bitcoin contracts%
\footnote{Bitcoin financial contracts:
  \href{https://www.colu.com/}{Colu},
  \href{http://coinspark.org/}{CoinSpark},
  \href{https://github.com/OpenAssets}{OpenAssets},
  \href{http://www.omnilayer.org/}{Omni},
  \href{https://www.smartbit.com.au/}{SmartBit},
  \href{https://bitpos.me/}{BitPos}.
}
totalize two thirds of the transactions published by the
373 Ethereum contracts in the same category. 

While both Bitcoin and Ethereum are mainly focussed on financial contracts,
we observe major differences about the other categories.
For instance, the Bitcoin contracts in the Notary category%
\footnote{Bitcoin notary contracts:
  \href{https://www.factom.com/}{Factom},
  \href{https://stampery.com/}{Stampery},
  \href{https://proofofexistence.com/}{Proof of Existence},
  \href{https://blocksign.com/}{Blocksign},
  \href{https://crypto-copyright.com/}{CryptoCopyright},
  \href{https://stampd.io/}{Stampd},
  \href{https://bitproof.io/}{BitProof},
  \href{https://github.com/thereal1024/ProveBit}{ProveBit},
  \href{https://remembr.io/}{Remembr},
  \href{https://originalmy.com/}{OriginalMy},
  \href{http://lapreuve.eu/explication.html}{LaPreuve},
  \href{http://digitalcurrency.unic.ac.cy/free-introductory-mooc/academic-certificates-on-the-blockchain/}{Nicosia},
  \href{http://www.chainpoint.org/}{Chainpoint},
  \href{http://diploma.report/}{Diploma},
  \href{https://monegraph.com/}{Monegraph},
  \href{https://blockai.com/}{Blockai},
  \href{https://www.ascribe.io}{Ascribe},
  \href{https://eternitywall.it/}{Eternity Wall},
  \href{https://github.com/blockstack/blockchain-id/wiki/Blockstore}{Blockstore}.
}
have an amount of transactions similar to that of the Financial category,
unlike in Ethereum.
The second most used category in Ethereum is Game. 
Although some games 
(\eg, lotteries~\cite{Andrychowicz14sp,Andrychowicz16cacm,Back13coin,Bentov14crypto} and poker~\cite{Kumaresan15ccs}) 
which run on Bitcoin have been proposed in the last few years,
the interest on them is still mainly academic,
and we have no experimental evidence that these contracts are used in practice.
Instead, the greater flexibility of the Ethereum programming language simplifies the development
of this kind of contracts (although with some quirks~\cite{Delmolino15bw} and limitations%
\footnote{%
  Although the Ethereum virtual machine is designed to be Turing-complete,
  in practice the limitations on the amount of gas which can be used to invoke contracts
  also limit the set of computable functions 
  (\eg, verifying checkmate exceeds the current gas limits of a transaction~\cite{Grau16chess}).
}). 

Note that in some cases there are not enough elements to categorise a contract.
This happens \eg, when the contract does not link to the project webpage, 
and there are neither comments in online forums nor in the contract sources.

\section{Design patterns for Ethereum smart contracts}
\label{sec:DesignPatterns}

In this~\namecref{sec:DesignPatterns} we study design patterns for Ethereum smart contracts.
To this purpose, we consider the sample of \ethcontracts contracts 
collected through the methodology described in~\Cref{sec:smartcontracts}.
By manually inspecting the Solidity source code of each of these contracts,
we identify some common design patterns.
We start in~\Cref{sec:patterns-list} by describing these patterns. 
Then, in~\Cref{sec:patterns-stats} we measure the usage of the patterns
in the various categories of contracts identified in~\Cref{sec:smartcontracts}.

  \subsection{Design patterns}
\label{sec:patterns-list}

\begin{description}

\item[Token.]  This pattern is used to distribute 
  some fungible goods (represented by tokens) to users. 
  Tokens can represent a wide variety of goods, 
  like \eg coins, shares, outcomes or tickets, or everything else
  which is transferable and countable.
  The implications of owning a token depend on the protocol and the
  use case for which the token has been issued. 
  Tokens can be used to track the ownership of physical properties 
  (\eg, \href{https://etherscan.io/address/0xe0b7927c4af23765cb51314a0e0521a9645f0e2a#code}{gold}~\cite{DGXWebsite}),
  or digital ones 
  (\eg, \href{https://etherscan.io/address/0x815a46107e5ee2291a76274dc879ce947a3f0850#code}{cryptocurrency}).
  Some crowdfunding systems issue tokens in exchange for donations
  (\eg, the \href{https://etherscan.io/address/0xfb6916095ca1df60bb79ce92ce3ea74c37c5d359#code}{Congress} contract).
  Tokens are also used to regulate user authorizations and identities. 
  For instance, the \href{https://etherscan.io/address/0xadc46ff5434910bd17b24ffb429e585223287d7f#code}{DVIP} 
  contract specifies 
  rights and term of services for owners of its tokens.
  To vote on the poll 
  \href{https://etherscan.io/address/0xdb6d68e1d8c3f69d32e2d83065492e502b4c67ba#code}{ETCSurvey},
  users must possess a suitable token. 
  Given the popularity of this pattern, its standardisation has been proposed~\cite{ERC20}. 
  Notably, the majority of the analysed Ethereum
  contracts which issue tokens already adhere to it.

\item[Authorization.] This pattern is used
  to restrict the execution of code according to the caller address.
  The majority of the analysed contracts check if the caller address is that of the contract owner,
  before performing critical operations 
  (\eg, sending ether, invoking
  \href{https://etherscan.io/address/0x3fccb426c33b1ae067115390354b968592348d05#code}{suicide}
  or 
  \href{https://etherscan.io/address/0x8b4aa759d83ec43efba755fc27923e4a581bccc1#code}{selfdestruct}).
  For instance, the owner of
  \href{https://etherscan.io/address/0xdc84953D7C6448e498Eb3C33ab0F815da5D13999#code}{Doubler}
  is authorized to move all funds to a new address \emph{at any time}
  (this may raise some concerns about the trustworthiness of the contract, 
  as a dishonest owner can easily steal money).
  \href{https://etherscan.io/address/0x684282178b1d61164febcf9609ca195bef9a33b5#code}{Corporation}
  checks addresses to ensure that every user can vote only once per poll.
  \href{https://etherscan.io/address/0x5A5eFF38DA95b0D58b6C616f2699168B480953C9#code}{CharlyLifeLog}
  uses a white-list of addresses to decide who can withdraw funds.

\item[Oracle.]  
  Some contracts may need to acquire data from outside the
  blockchain, \eg from a website, to determine the winner of a bet.
  The Ethereum language does not allow contracts to query external sites:
  otherwise, the determinism of computations would be broken,
  as different nodes could receive different results for the same query.
  Oracles are the interface between contracts and the outside.
  Technically, they are just contracts, 
  and as such their state can be updated by sending them transactions.
  In practice, instead of querying an external service, a contract queries an oracle;
  and when the external service needs to update its data, 
  it sends a suitable transaction to the oracle.
  Since the oracle is a contract, it can be queried from other contracts 
  without consistency issues.
  One of the most common oracles is Oraclize%
  \footnote{\url{http://www.oraclize.it/}}:
  in our sample, it is used by almost all the contracts which resort to oracles.

\item[Randomness.]  Dealing with randomness is not a trivial task in Ethereum.
  Since contract execution must be deterministic, 
  all the nodes must obtain the same value when asking for a random number: 
  this struggles with the randomness requirements wished.
  To address this issue, several contracts 
  (\eg, \href{https://etherscan.io/address/0x76bc9e61a1904b82cbf70d1fd9c0f8a120483bbb#code}{Slot})
  query oracles that generate these values off-chain. 
  Others
  (\eg, \href{https://etherscan.io/address/0x302fE87B56330BE266599FAB2A54747299B5aC5B#code}{Lottery}) 
  try to generate the numbers locally, by using values
  not predictable \emph{a priori}, as the hash of a block not yet created.
  However, these techniques are not generally considered secure~\cite{ABC17post}.
	
\item[Poll.] Polls allows users to vote on some question.
  Often this is a side feature in a more complex scenario.
  For instance, in the  
  \href{https://etherscan.io/address/0x2AB9f67A27f606272189b307052694D3a2B158bA#code}{Dice} game,
  when a certain state is reached, the owner issues a poll
  to decide whether an emergency withdrawal is needed.
  To determine who can vote and to keep track of the votes, 
  polls can use tokens, or they can check the voters' addresses.

\item[Time constraint.]  Many contracts implement time constraints, 
  \eg to specify when an action is permitted.
  For instance,
  \href{https://etherscan.io/address/0x9828f591b21ee4ad4fd803fc7339588cb83a6b84#code}{BirthdayGift}
  allows users to collect funds, which will be redeemable only after their birthday.
  In notary contracts, time constraints are used to prove that a
  document is owned from a certain date. 
  In game contracts,
  \eg \href{https://etherscan.io/address/0x302fE87B56330BE266599FAB2A54747299B5aC5B#code}{Lottery},
  time constraints mark the stages of the game.

\item[Termination.] Since the blockchain is immutable, a
  contract cannot be deleted when its use has come to an end. 
  Hence, developers must forethink a way to disable it, 
  so that it is still present but unresponsive. 
  This can be done manually, by inserting ad-hoc code in the contract, 
  or automatically, calling \code{selfdestruct} or \code{suicide}. 
  Usually, only the contract owner is authorized to terminate a contract 
  (\eg, as in \href{https://etherscan.io/address/0xe941e5d4a66123dc74886699544fbbb942f1887a#code}{SimpleCoinFlipGame}).
	
\item[Math.] Contracts using this pattern encode the logic 
  which guards the execution of some critical operations.
  For instance, 
  \href{https://etherscan.io/address/0x54bda709fed875224eae569bb6817d96ef7ed9ad#code}{Badge}
  implements a method named \code{subtractSafely} 
  to avoid subtracting a value from a balance 
  when there are not enough funds in an account.
	
\item[Fork check.] The Ethereum blockchain has been forked four times,
  starting from July 20th, 2016,
  when a fork was performed to contrast the effect of the DAO attack~\cite{TheDaoHardFork}. 
  To know whether or not the fork took place, 
  some contracts inspect the final balance of the DAO. 
  Other contracts use this check
  to detect whether they are running on the main chain or on the fork, 
  performing different actions in the two cases.
  \href{https://etherscan.io/address/0x2bd2326c993dfaef84f696526064ff22eba5b362#code}{AmIOnTheFork}
  is a library contract that can be used to distinguish 
  the main chain from the forked one.

\end{description}

  \subsection{Quantifying the usage of design patterns by category}
\label{sec:patterns-stats}

\begin{table}[t]
  \begin{center}
    \hspace*{0em}
    \small
    \scalebox{0.9}{%
      \begin{tabular}{|l|c|c|c|c|c|c|c|c|c||c|}
        \hline
        \multirow{2}{5em}{}& \textbf{Token} & \textbf{Auth.} & \textbf{Oracle} & \textbf{Random.} & \textbf{Poll} & \textbf{Time} & \textbf{Termin.} & \textbf{Fork} & \textbf{Math} & \textbf{None} \\
        \hline
        \textbf{Financial} & 24-51 & 51-39 & 2-15 & 1-2 & 5-29 & 23-31 & 14-30 & 8-69 & 4-47 & 29-66 \\ 
        \textbf{Notary} & 13-6 & 52-9 & 1-2 & 0-0 & 8-9 & 20-6 & 29-13 & 0-0 & 1-3 & 30-15 \\ 
        \textbf{Game} & 3-3 & 84-27 & 25-74 & 72-93 & 25-57 & 73-43 & 21-19 & 1-3 & 2-9 & 1-1 \\ 
        \textbf{Wallet} & 18-2 & 100-3 & 0-0 & 0-0 & 0-0 & 94-6 & 100-10 & 0-0 & 12-6 & 0-0 \\ 
        \textbf{Library} & 0-0 & 31-2 & 0-0 & 14-3 & 0-0 & 24-3 & 24-4 & 34-24 & 21-19 & 17-3 \\ 
        \hline\hline
        \textbf{Unclassified} & 43-39 & 66-21 & 3-9 & 1-1 & 3-6 & 18-10 & 28-25 & 28-25 & 1-5 & 15-15 \\ 
        \textbf{\emph{Total}} & {\em 21-100} & {\em 61-100} & {\em 7-100} & {\em 15-100} & {\em 9-100} & {\em 33-100} & {\em 22-100} & {\em 5-100} & {\em 4-100} & {\em 20-100} \\ 
        \hline
      \end{tabular}
    } 
  \end{center}
  \caption{Relations between design patterns and contract categories.
    A pair $(p,q)$ at row $i$ and column $j$ means that
    $p\%$ of the contracts in category $i$ use the pattern of column $j$,
    and $q\%$ of  contracts with pattern $j$ belong to category $i$.}
  \label{tab:patterns}
\end{table}

We now study how the design patterns identified in
\Cref{sec:patterns-list} are used in  smart contracts.
Out of the \ethcontracts analysed contracts, 648 use at least one of the 9 patterns
presented, for a grand total of 1427 occurrences of usage.

\Cref{tab:patterns} shows the correlation between
the usage of design patterns and contract categories,
as defined in~\Cref{sec:smartcontracts}.  
A cell at row $i$ and column $j$ shows a pair of values:
the first value is the percentage of contracts of category $i$ that use the pattern of column $j$;
the second one is the percentage of contracts with pattern $j$ which belongs to category $i$.
So, for instance,
24\% of the contracts in the financial category use the token pattern,
and 51\% of all the contracts with the token pattern are financial ones.

We observe that \textit{token}, \textit{authorization}, \textit{time constraint},
and \textit{termination} are generally the most used patterns.
Some patterns are spread across several categories 
(\eg, \emph{termination} and \textit{time constraint}), 
while others are mainly adopted only in one. 
For instance, \emph{oracle} and \emph{randomness} patterns are peculiar of game contracts,
while the \emph{token} pattern is mostly used in financial contracts.
Although \textit{math} is the less used, it appears in each category. 
Some contracts do not use any pattern (29\% of financial and 30\% of notary); 
almost all the contracts in game and wallet categories uses at least one.
Further, only 15\% of all the unclassified contracts do no use any pattern at all.

The most frequent patterns in financial contracts are
\textit{token} (24\%), \textit{authorization} (51\%), 
and \textit{time constraint} (23\%).
Due to the presence of contracts which implement assets and crowdfunding services, 
we have that half of contracts using \textit{token} and \textit{math} patterns belong 
to the financial category. For instance, these  services use \textit{token} for 
representing goods or developing polls. 
Moreover, a great 69\% of contracts that use the \textit{fork check}
pattern is financial. 
This is caused by the necessity of knowing the
branch of the fork before deciding to move funds.
Finally, several financial applications (29\%) perform simple
operations (\eg sending a payment) without using any of our described patterns.

The \textit{authorization} pattern is used in many notary contracts 
to ensure that only the owner of a document can add or modify its data, 
in order to avoid tampering.
Most gambling games involve players who pay fees to join the game,
and rewards that can be collected by the winner of the game.
The \emph{authorization} pattern is used to let the owner to be the only one
able to redeem participants' fees or to perform administrative
operations, and to let the winner withdraw his reward.
The \emph{time constraint} pattern is used to distinguish the different phases of the game.
For instance, within a specific time interval players can join the game and/or bet;
then, bets are over, and the game determines a winner. 
To choose the winner, some gambling games resort to
random numbers, which are often generated through an oracle. 
Indeed, 25\% of games use the \textit{oracle} pattern, 
and the pattern itself is used 74\% of cases by a game contract. 
Since \emph{all} game contracts invoking an \emph{oracle} (25\%) ask for random values, 
and since 72\% of contracts use the \emph{random} pattern, 
we can deduce that 47\% of them generate random numbers without resorting to oracles.

Notably, 100\% of wallet contracts adopt both \textit{authorization}
and \textit{termination} design patterns. 
A high 94\% also uses \textit{time constraint}. 
On the contrary, \textit{oracle}, \textit{poll}, and \textit{randomness} patterns 
are of little use when developing a wallet, 
while \textit{math} is sometimes used for securing operations on the balance.

\section{Conclusions}

We have analysed the usage of smart contracts from various perspectives.
In~\Cref{sec:platforms} we have examined a sample of 6 platforms for smart contracts,
pinpointing some crucial technical differences between them. 
For the two most prominent platforms --- Bitcoin and Ethereum ---
we have studied a sample of \totcontracts contracts,
categorizing each of them by its application domain,
and measuring the relevance of each of these categories
(\Cref{sec:smartcontracts}).
The availability of source code for Ethereum contracts has allowed us to analyse
the most common design patterns adopted when writing smart contracts (\Cref{sec:DesignPatterns}). 

We believe that this survey may provide valuable information 
to developers of new, domain-specific languages for smart contracts. 
In particular, measuring what are the most common use cases 
allows to understand which domains deserve more investments. 
Furthermore, our study of the correlation between design patterns and 
application domains can be exploited to drive the correct choice of 
programming primitives of domain-specific languages for smart contracts.

Due to the mixed flavour of our analysis, which compares differents
platforms and studies how smart contracts are interpreted on each them,
our work relates to various topics.
The work~\cite{Marino16ruleml} proposes design patterns for altering and 
undoing of smart contracts;
so far, our analysis in~\Cref{sec:patterns-stats} has not still found instances of these patterns
in Ethereum.
Among the works which study blockchain technologies,
\cite{Anderson16corr} compares four blockchains, with a special focus on the Ethereum one; 
\cite{Lamela16iacr} examines a larger set of blockchains,
including also some which does not fit the criteria we have used in our methodology 
(\eg, RootStock and Tezos).
Many works on Bitcoin perform empirical analyses of its blockchain.
For instance, 
\cite{reid2013analysis,ron2013quantitative} study users deanonymization,
\cite{moser2015trends} measures transactions fees,
and~\cite{baqerstressing} analyses Denial-of-Service attacks on Bitcoin.  
The work~\cite{Glaser14ecis} investigates whether Bitcoin users are
interested more on digital currencies as asset or as currency, with
the aim of detecting the most popular use cases of Bitcoin contracts,
similarly to what we have done in~\Cref{sec:smartContracts:quantitative}.
Our classification of Bitcoin protocols based on \opreturn transactions
is inspired from~\cite{Bartoletti16opreturn}, which 
also measures the space consumption and temporal trend of \opreturn transactions.

Recently, 
some authors have started to analyse the security of Ethereum smart contracts: 
among these, 
\cite{ABC17post} surveys vulnerabilities and attacks,
while~\cite{Luu16ccs} and~\cite{Bhargavan16solidether}
propose analysis techniques to detect them.
Our study on design patterns for Ethereum smart contracts could
help to improve these techniques,
by targeting contracts with specific programming patterns.

  \subsubsection*{Acknowledgments.} 

This work is partially supported by Aut.\ Reg.\ of Sardinia project P.I.A.\ 2013 ``NOMAD''.

\bibliographystyle{splncs03}
\bibliography{main}

\begin{thebibliography}{10}
\providecommand{\url}[1]{\texttt{#1}}
\providecommand{\urlprefix}{URL }

\bibitem{BitcoinContract}
Bitcoin contract, \url{https://en.bitcoin.it/wiki/Contract}. Last accessed
  2017/01/14

\bibitem{BitcoinOpReturnDescription}
Bitcoin {OP\_RETURN} wiki page, \url{https://en.bitcoin.it/wiki/OP_RETURN}.
  Last accessed 2017/01/14

\bibitem{DGXWebsite}
Dgx website, \url{https://www.dgx.io/}. Last accessed 2017/01/14

\bibitem{TheDaoHardFork}
Ethereum hard fork 20 july 2016,
  \url{https://blog.ethereum.org/2016/07/20/hard-fork-completed/}. Last
  accessed 2017/01/14

\bibitem{ERC20}
Ethereum request for comment 20,
  \url{https://github.com/ethereum/wiki/wiki/Standardized_Contract_APIs}. Last
  accessed 2017/01/14

\bibitem{Lisk}
Lisk, \url{https://lisk.io/}. Last accessed 2017/01/14

\bibitem{MakingSenseContracts}
Making sense of blockchain smart contracts,
  \url{http://www.coindesk.com/making-sense-smart-contracts/}. Last accessed
  2017/01/14

\bibitem{Monax}
Monax, \url{https://monax.io/}. Last accessed 2017/01/14

\bibitem{MultiChainGoodBadLazy}
Smart contracts: The good, the bad and the lazy,
  \url{http://www.multichain.com/blog/2015/11/smart-contracts-good-bad-lazy/}.
  Last accessed 2017/01/14

\bibitem{Stellar}
Stellar, \url{https://www.stellar.org/}. Last accessed 2017/01/14

\bibitem{SCP}
The {Stellar} consensus protocol,
  \url{https://www.stellar.org/papers/stellar-consensus-protocol.pdf}. Last
  accessed 2017/01/14

\bibitem{VitalikThinkingSecurity}
Thinking about smart contract security,
  \url{https://blog.ethereum.org/2016/06/19/thinking-smart-contract-security/}.
  Last accessed 2017/01/14

\bibitem{DAO}
Understanding the {DAO} attack,
  \url{http://www.coindesk.com/understanding-dao-hack-journalists/}. Last
  accessed 2017/01/14

\bibitem{ENSanotherbug}
Another bug in the ens, you can win with an unlimited high bid without paying
  for it (2017),
  \url{https://www.reddit.com/r/ethereum/comments/5zctus/another_bug_in_the_ens_you_can_win_with_an/}.
  Last accessed 2017/03/17

\bibitem{Anderson16corr}
Anderson, L., Holz, R., Ponomarev, A., Rimba, P., Weber, I.: New kids on the
  block: an analysis of modern blockchains. CoRR  abs/1606.06530 (2016)

\bibitem{Andrychowicz14sp}
Andrychowicz, M., Dziembowski, S., Malinowski, D., Mazurek, L.: Secure
  multiparty computations on {Bitcoin}. In: {IEEE} \mbox{S \& P}. pp. 443--458
  (2014)

\bibitem{Andrychowicz16cacm}
Andrychowicz, M., Dziembowski, S., Malinowski, D., Mazurek, L.: Secure
  multiparty computations on {Bitcoin}. Commun. {ACM}  59(4),  76--84 (2016),
  \url{http://doi.acm.org/10.1145/2896386}

\bibitem{ABC17post}
Atzei, N., Bartoletti, M., Cimoli, T.: A survey of attacks on {Ethereum} smart
  contracts. Cryptology ePrint Archive, Report 2016/1007 (2016),
  \url{http://eprint.iacr.org/2016/1007}

\bibitem{Back13coin}
Back, A., Bentov, I.: Note on fair coin toss via {Bitcoin}.
  \url{http://www.cs.technion.ac.il/~idddo/cointossBitcoin.pdf} (2013)

\bibitem{Banasik16esorics}
Banasik, W., Dziembowski, S., Malinowski, D.: Efficient zero-knowledge
  contingent payments in cryptocurrencies without scripts. In: {ESORICS}. pp.
  261--280 (2016)

\bibitem{baqerstressing}
Baqer, K., Huang, D.Y., McCoy, D., Weaver, N.: Stressing out: Bitcoin ``stress
  testing''. In: {Bitcoin} Workshop. pp. 3--18 (2016)

\bibitem{Bartoletti17ponzi}
Bartoletti, M., Carta, S., Cimoli, T., Saia, R.: Dissecting {Ponzi} schemes on
  {Ethereum}: identification, analysis, and impact. CoRR  abs/1703.03779
  (2017), \url{https://arxiv.org/abs/1703.03779}

\bibitem{Bartoletti16opreturn}
Bartoletti, M., Pompianu, L.: An analysis of {Bitcoin} {OP\_RETURN} metadata.
  CoRR  abs/1702.01024 (2016), \url{https://arxiv.org/abs/1702.01024}, to
  appear in {Bitcoin} {Workshop} 2017

\bibitem{Bentov14crypto}
Bentov, I., Kumaresan, R.: How to use {Bitcoin} to design fair protocols. In:
  {CRYPTO}. pp. 421--439 (2014)

\bibitem{Bhargavan16solidether}
Bhargavan, K., Delignat-Lavaud, A., Fournet, C., Gollamudi, A., Gonthier, G.,
  Kobeissi, N., Rastogi, A., Sibut-Pinote, T., Swamy, N., Zanella-Beguelin, S.:
  Formal verification of smart contracts. In: {PLAS} (2016)

\bibitem{Bonneau15ieeesp}
Bonneau, J., Miller, A., Clark, J., Narayanan, A., Kroll, J.A., Felten, E.W.:
  {SoK}: Research perspectives and challenges for {Bitcoin} and
  cryptocurrencies. In: {IEEE} \mbox{S \& P}. pp. 104--121 (2015)

\bibitem{Brown16corda}
Brown, R.G., Carlyle, J., Grigg, I., Hearn, M.: {Corda}: An introduction.
  \url{http://r3cev.com/s/corda-introductory-whitepaper-final.pdf} (2016)

\bibitem{ethereum}
Buterin, V.: {Ethereum}: a next generation smart contract and decentralized
  application platform. \url{https://github.com/ethereum/wiki/wiki/White-Paper}
  (2013)

\bibitem{Churyumov16byteball}
Churyumov, A.: {Byteball}: a decentralized system for transfer of value.
  \url{https://byteball.org/Byteball.pdf} (2016)

\bibitem{Clack16corr}
Clack, C.D., Bakshi, V.A., Braine, L.: Smart contract templates: foundations,
  design landscape and research directions. CoRR  abs/1608.00771 (2016)

\bibitem{Delmolino15bw}
Delmolino, K., Arnett, M., Miller, A., Kosba, A., Shi, E.: Step by step towards
  creating a safe smart contract: Lessons and insights from a cryptocurrency
  lab. In: {Bitcoin} Workshop (2016)

\bibitem{counterparty}
Dermody, R., Krellenstein, A., Slama, O., Wagner, E.: {Counterparty}: Protocol
  specification (2014),
  \url{http://counterparty.io/docs/protocol_specification/}. Last accessed
  2017/01/14

\bibitem{Frantz16ecas}
Frantz, C.K., Nowostawski, M.: From institutions to code: towards automated
  generation of smart contracts. In: Workshop on Engineering Collective
  Adaptive Systems ({eCAS}) (2016)

\bibitem{Glaser14ecis}
Glaser, F., Zimmermann, K., Haferkorn, M., Weber, M.C.: {Bitcoin} - asset or
  currency? revealing users' hidden intentions. In: European Conference on
  Information Systems ({ECIS}) (2014)

\bibitem{Grau16chess}
Grau, P.: Lessons learned from making a chess game for {Ethereum} (2016),
  \url{https://medium.com/@graycoding/lessons-learned-from-making-a-chess-game-for-ethereum-6917c01178b6#.fwtdwly6e}.
  Last accessed 2017/01/14

\bibitem{Kumaresan15ccs}
Kumaresan, R., Moran, T., Bentov, I.: How to use {Bitcoin} to play
  decentralized poker. In: {ACM} {CCS}. pp. 195--206 (2015)

\bibitem{Luu16ccs}
Luu, L., Chu, D.H., Olickel, H., Saxena, P., Hobor, A.: Making smart contracts
  smarter. In: {ACM} {CCS} (2016), \url{http://eprint.iacr.org/2016/633}

\bibitem{Marino16ruleml}
Marino, B., Juels, A.: Setting standards for altering and undoing smart
  contracts. In: {RuleML}. pp. 151--166 (2016)

\bibitem{moser2015trends}
M{\"o}ser, M., B{\"o}hme, R.: Trends, tips, tolls: A longitudinal study of
  bitcoin transaction fees. In: Financial Cryptography and Data Security. pp.
  19--33 (2015)

\bibitem{bitcoin}
Nakamoto, S.: {Bitcoin}: a peer-to-peer electronic cash system.
  \url{https://bitcoin.org/bitcoin.pdf} (2008)

\bibitem{Japan16report}
{Nomura Research Institute}: Survey on blockchain technologies and related
  services, \url{http://www.meti.go.jp/english/press/2016/pdf/0531\_01f.pdf}

\bibitem{Popejoy16kadena}
Popejoy, S.: The {Pact} smart contract language. \url{http://kadena.io/pact}
  (2016)

\bibitem{reid2013analysis}
Reid, F., Harrigan, M.: An analysis of anonymity in the {Bitcoin} system. In:
  Security and privacy in social networks, pp. 197--223. Springer (2013)

\bibitem{ron2013quantitative}
Ron, D., Shamir, A.: Quantitative analysis of the full {Bitcoin} transaction
  graph. In: Financial Cryptography and Data Security. pp. 6--24. Springer
  (2013)

\bibitem{Lamela16iacr}
Seijas, P.L., Thompson, S., McAdams, D.: Scripting smart contracts for
  distributed ledger technology. Cryptology ePrint Archive, Report 2016/1156
  (2016), \url{http://eprint.iacr.org/2016/1156}

\bibitem{Sompolinsky15fc}
Sompolinsky, Y., Zohar, A.: Secure high-rate transaction processing in bitcoin.
  In: Financial Cryptography and Data Security. pp. 507--527 (2015)

\bibitem{Szabo97firstmonday}
Szabo, N.: Formalizing and securing relationships on public networks. First
  Monday  2(9) (1997),
  \url{http://firstmonday.org/htbin/cgiwrap/bin/ojs/index.php/fm/article/view/548}

\bibitem{UK16report}
{{UK} {Government} {Chief} {Scientific} {Adviser}}: Distributed ledger
  technology: beyond block chain,
  \url{https://www.gov.uk/government/uploads/system/uploads/attachment\_data/file/492972/gs-16-1-distributed-ledger-technology.pdf}

\bibitem{ethereumyellowpaper}
Wood, G.: {Ethereum}: a secure decentralised generalised transaction ledger.
  \url{gavwood.com/paper.pdf} (2014)

\end{thebibliography}

\end{document}